\documentclass[pre,superscriptaddress,twocolumn,a4paper,showpacs]{revtex4-1}
\usepackage{graphics}
\usepackage{amssymb}
\usepackage{wasysym}
\usepackage{latexsym}
\usepackage{graphicx}
\usepackage{epsfig}
\usepackage{subfigure}
\begin{document}

\title{On the efficacy of the wisdom of crowds to forecast economic indicators}

\author{Nilton S. Siqueira Neto}
\affiliation{Instituto de F\'{\i}sica de S\~ao Carlos,
  Universidade de S\~ao Paulo,
  Caixa Postal 369, 13560-970 S\~ao Carlos, S\~ao Paulo, Brazil}

\author{Jos\'e F.  Fontanari}
\affiliation{Instituto de F\'{\i}sica de S\~ao Carlos,
  Universidade de S\~ao Paulo,
  Caixa Postal 369, 13560-970 S\~ao Carlos, S\~ao Paulo, Brazil}

\begin{abstract}
The  interest in the wisdom of crowds stems mainly from the possibility of combining independent  forecasts from experts  in the hope that  many expert minds are   better than a few. Hence the  relevant subject of study nowadays is the \textit{Vox Expertorum} rather than Galton's original \textit{Vox Populi}. Here we use the  Federal Reserve Bank of Philadelphia's   Survey of Professional Forecasters  to analyze  $15455$   forecasting contests to predict a variety of economic indicators. We find that the median has  advantages over the mean as a method to combine the experts' estimates:  the odds that the crowd beats all participants of a forecasting contest is $0.015$ when the aggregation is given by the mean   and $0.026$ when it  is given by the median.  In addition, the median is always guaranteed to beat the majority of the participants, whereas the mean beats that majority in 67 percent of the forecasts only. Both aggregation methods yield  a $20$ percent error on the average, which  must be  contrasted with  the  $15$ percent error of the contests' winners. A   standard time series forecasting algorithm, the  ARIMA  model, yields   a  $31$ percent error on the average.
However, since the expected error of a randomly selected forecaster is about $22$ percent, our conclusion is that selective attention is the most likely explanation for the mysterious high accuracy of the crowd reported in the literature. 
\end{abstract}

\maketitle

%
\section{Introduction} \label{sec:intro}
 %

The phenomenon known as  Vox Populi or  wisdom of crowds  is more than a century old, being brought to light by Galton's 1907  study  of  a contest to guess  the weight of an ox  at the West of England Fat Stock and Poultry Exhibition in Plymouth \cite{Galton1907}  (see  \cite{Wallis2014} for a brief historical account).  Galton  used the median of the participants' estimates as the collective estimate and  found that   it  overestimated the weight of the ox by less than 1 percent of the true weight \cite{Galton1907,Wallis2014}.  The wisdom of crowds has been brought to the attention of a wider audience by a series of business bestsellers  published in the early 2000s \cite{Surowiecki2004,Sunstein2006,Page2007}, but  the remarkable accuracy of the estimate  that results from the aggregation  of many independent individuals' judgments is still somewhat of a mystery today.  

A difficulty  to address the wisdom of crowds is that it seems to have distinct meanings for  different researchers.  On the one hand, some  researchers view this phenomenon as  the finding that the crowd can solve problems better than most individuals in it, including experts \cite{Surowiecki2004}.  In this perspective, the actual accuracy of the crowd estimate is not relevant: what matters is that its estimate is better than the estimate of most of the participants of the forecasting contest. In other words, the crowd is viewed as another forecaster and the  focus is on its rank among the participants. On the other hand, some researchers stress the accuracy of the crowd estimate, despite the large dispersal of the individuals'  judgments, without much regard to its rank  \cite{Lorenz2011}.     This latter perspective seems more fit to describe Galton's original reaction to the  surprising trust-worthiness of the popular judgment \cite{Galton1907}. Here  we consider extensively these two perspectives.

It is interesting that both views of the wisdom of crowds  are trivially vindicated  depending on the method we choose to aggregate the  independent individuals' judgments. For instance, if,  on the one hand,    we adopt   Galton's suggestion and choose the median of the individuals' judgments as the collective estimate, then it follows that the crowd will always beat at least half of the individuals. In fact, assuming that the true value of  the ox weight is to the right of the median then the median is closer to the true value than all  the individuals' estimates that are to the left of the median, which accounts to half of the total estimates. A similar logic applies to the case the true value is to the left of the median. On the other hand, if we adopt the mean of the individuals' judgments as the collective estimate we can easily show that the  error of the collective estimate can never be greater  than the average individual  error, which equals the expected error of the estimate of a randomly selected  participant \cite{Page2007}.   As an aside, we note that in  Galton's ox-weighing judgment the choice of the  mean as the collective estimate yields zero error \cite{Wallis2014}.

A  natural explanation for the wisdom of crowds  calls upon  the  well-known fact that the  combination of unbiased and independent
estimates   guarantees  the accuracy of the statistical average, provided  the number of estimates is large  \cite{Bates1969}. In other words, if the judgments made by numerous different people scatter symmetrically  around the truth, then the collective estimate is likely to be very accurate.  Somewhat surprisingly, this explanation is rather popular and claims that ``the wisdom of crowds effect works if estimation errors of individuals are large but unbiased such that they cancel each other out'' abound in the literature \cite{Lorenz2011}. It is of course unarguably true that if the  individuals' judgments are unbiased then the wisdom of crowds effect will hold.  The trouble with this explanation is  the assumption that  the independent individuals' judgments are unbiased, that is, that their means coincide with the true value of the unknown quantity  so that there are no systematic errors on the individuals' estimates. The typical data do not support this assumption. In fact,  the pronounced asymmetry of the distribution of individuals' judgments  has not eluded Galton's attention, who even tried to fit it with a two-piece distribution \cite{Wallis2014}.  The nonzero  skewness of the distribution of individual judgments offers  strong evidence for  biases on the  individuals' judgments \cite{Nash2014,Nash2017}. In addition, since  there is no meaningful correlation between the asymmetry of the distribution of the individuals' judgments and the collective estimation error \cite{Reia2021}, the  cancellation  of the participants' estimation errors  cannot be the general  explanation for the wisdom of crowds.

Another difficulty to get to the bottom of the wisdom of crowds  is that much of the evidence in favor of its existence has relied on anecdotes, such as the celebrated Galton's ox-weighing contest. This may lead  to the  suspicion that selective attention is at play, i.e.,  that prominence is given to the successful  outcomes only. For instance,  the analysis of three wisdom of crowds experiments, viz., the  estimate of the  number of candies in a jar,  the estimate of the length of a paper strip  and the estimate of the number of pages of a book, in which the crowd estimate was given by the  arithmetic mean of the participants' judgments, resulted in $16.5$,  $1.8$,  and  $28.4$ percent   errors, respectively \cite{Nobre2020}. Nevertheless, one hears mostly of the high accuracy of the crowd estimates \cite{Surowiecki2004,Sunstein2006,Page2007}.  As a matter of fact, it is not possible to assess the accuracy of an estimate without considering some standard of comparison and in this paper we offer some  proposals in that direction.  Of course, a simple and  natural comparison standard is the prediction of the winner of a forecasting contest.  In particular,  for the  three wisdom of crowds experiments  mentioned above, the percent   errors of the winners' predictions were  $0.9$, $0.4$ and $2$, respectively  \cite{Nobre2020}. 

To avoid the selective attention fallacy and aim at  statistically meaningful  conclusions we need to consider  very many  independent wisdom of crowds  experiments or forecasting contests. Each experiment is fully characterized by the distribution of   individuals' estimates.  Here we accomplish this by using  forecasts of economic indicators that  are publicly available   in the Federal Reserve Bank of Philadelphia's  Survey of Professional Forecasters  (FRBP-SPF) database \cite{FRBP}.  In particular, we  consider quarterly projections  of $20$  economic indicators, which amounts to  $15455$  independent forecasting contests.

In this paper  we expand on a previous study of the FRBP-SPF database  \cite{Reia2021} by almost doubling the number of forecasts considered and  by comparing two methods to combine the individuals' estimates, viz.,  the mean and the median. In addition, we assess the accuracy of the crowd's performance by comparing it with  the standard time series forecasting algorithm, the  ARIMA (AutoRegressive Integrated Moving Average) model,  as well as with the winners of the forecasting contests. Overall we find that the median is superior to the mean as a means to combine the individuals' judgments, but both choices yield  a $20$ percent error on the average.     This is a great performance when compared with  ARIMA's that yields a $31$ percent error on the average and it does not  fare badly when  compared with the $15$ percent error of the contests' winners.   
However, since the expected error of a randomly selected forecaster is about $22$ percent, our conclusion is that selective attention is the most likely explanation for the mysterious accuracy of the crowd. This conclusion is strengthened by the debunking of the popular explanation that the participants' errors cancel each other out resulting in the high accuracy of the crowd estimate.

\section{The FRBP-SPF  database}\label{sec:data} 

The Federal Reserve Bank of Philadelphia's Survey of Professional Forecasters (FRBP-SPF)  offers quarterly projections for five quarters of a variety of economic indicators since 1968 \cite{FRBP} (see \cite{Clements2022} for a recent review on the statistical literature based on that database).  This survey  provides forecasts of the current quarter (i.e.,
of the quarter in which the survey is held) and of each of the next four quarters.   For instance,  consider  a survey of a particular variable, say NGDP, held in the first quarter of 2006 (or 2006:Q1 in the notation of the documentation of the FRBP-SPF database). The last known historical quarter at that date is 2005:Q4 and the quarterly observation dates forecast are 2006:Q1 (i.e., the current quarter),  2006:Q2,  2006:Q3,  2006:Q4 and  2007:Q1.
We focus on the surveys of the 20 economic indicators, viz., NGDP, PGDP, CPROF, UNEMP, EMP, INDPROD, HOUSING, TBILL, TBOND, RGDP, RCONSUM, RNRESIN, RRESINV, RFEDGOV, RSLGOV, REXPORT, CPI, CORE\-CPI, PCE, and COREPCE  from the date they entered the FRBP-SPF database until  2020:Q4, which means we need  the quarterly historical values of those variables until 2021:Q4.
The  variables NGDP, PGDP, CPROF, UNEMP, INDPROD, HOUSING, and RGDP were the first to enter the FRBP-SPF database  (1968:Q4) and CORECPI was the last (2007:Q1).   
All forecasters are select  economists and  the mean number of  participants for survey is $\langle n \rangle = 35$, whereas  the minimum number is $n_{min}= 8$ and the maximum  $n_{max}= 87$.

At this stage we can define a forecasting contest (or a wisdom of crowds  experiment) in the framework of the  FRBP-SPF database more precisely. We recall that, similarly to Galton's ox-weighing contest, a forecasting contest involves a group of individuals making independent judgments about the  value of an unknown quantity. For instance, consider the forecast of the  variable NGDP for 2021:Q4.  There are five  distinct forecasting contests (one for each forecast horizon) to predict the value of that variable, viz., the surveys held at  2021:Q4,  2021:Q3,  2021:Q2, 2021:Q2 and 2020:Q4. Hence each survey of a particular economic variable can be seen as a forecasting contest.  We note that  not necessarily the same economists participate  of all those surveys,  since they held at different dates.

Since the forecasters are unaware of each other's predictions, it is not too far-fetched to assume that their predictions are independent, which seems a  necessary condition for the wisdom of crowds to work \cite{Lorenz2011,King2011}.  We clump together the 20  variables listed above as well as the 5 forecast horizons  to form a  large ensemble of  forecasting contests.  Since we need to use percentage errors to compare the forecasters' performances in those different contests, we have eliminated the  contests for which the true (historical) value of the variable is zero.  Fortunately,  only the variables CPI and PCE have historical values equal  zero  and those happened for  2013:Q2 only.  As there are five forecasting contests  associated to each of those values, namely,  the contests held at 2013:Q2, 2013:Q1, 2012:Q4, 2012:Q3, and 2012:Q2,  we have eliminated solely 10   forecasting contests in  total.
Overall the total number of forecasting contests  in our study is  $15455$. This amount contrasts with previous (mostly anecdotal) analyses  that considered only a few forecasting contests.

 As pointed out,  the FRBP database  considered here consists of predictions of expert economists, which contrasts with the common,  but essentially  incorrect, view of the wisdom of crowds as the aggregation of predictions of ordinary people.  Of course,  the wisdom of crowds  as a method of information aggregation depends on the presence of an expressive number of experts in the crowd as  nothing good can come from  the aggregation of random information \cite{Sunstein2006}.  In fact, the interest  of business-minded researchers  in the wisdom of crowds stems from the possibility of combining the forecasts of experts  in the hope that  many expert minds are   better than a few.  It is interesting that just after the publication of  Galton's paper on the ox-weighing contest, it was pointed out that Galton had reported the efficacy of \textit{Vox Expertorum} rather than of \textit{Vox Populi},  as  the participants of the contest were most likely butchers and farmers whose livelihood depended on their ability to judge the weight of farm animals before trading \cite{Coste1907}. Moreover, the 6 pence    tickets probably deterred the entrance of dilettantes in the contest.

%
\section{Mean \textit{versus} Median }\label{sec:MM} 
%
The first question we explore is whether there is an advantage  of choosing the mean over the median (and vice versa)   to combine the independent participants' judgments.  We recall that in the ox-weighing contest, the median overestimated the weight by 0.8 percent whereas the mean yielded zero error \cite{Wallis2014}.  

Let $g_i$ be the estimate of some unknown quantity (e.g.,  the weight of the ox in Galton's contest)  by individual $i=1, \ldots, n$, so  
the arithmetic mean of the individuals' estimates is 
\begin{equation}\label{mean}
  \langle g \rangle_n =  \frac{1}{n} \sum_{i=1}^n g_i.
\end{equation}
If we denote by $G$  the true value of the unknown quantity, then the collective error resulting from the use of the arithmetic mean to aggregate the individuals' estimates is
\begin{equation}\label{gamma_1}
 \gamma_{mean} =  | \langle g \rangle_n - G | .
\end{equation}
Defining the average quadratic individual error as
\begin{equation}\label{eps}
\epsilon_{quad} = \frac{1}{n} \sum_{i=1}^n \left ( g_i - G \right )^2,
\end{equation}
and the diversity of the estimates as
\begin{equation}\label{delta}
\delta = \frac{1}{n} \sum_{i=1}^n \left ( g_i  - \langle g \rangle_n  \right )^2,
\end{equation}
we obtain the  identity
\begin{equation}\label{DPT}
\gamma_{mean}^2 =  \epsilon_{quad} - \delta, 
\end{equation}
which is Page's diversity prediction theorem \cite{Page2007}. This theorem   asserts  that, for a given forecasting contest, the  (quadratic) collective error  equals the average (quadratic) individual error minus the prediction diversity.  This result is sometimes viewed as indication that the increase of the prediction diversity $\delta$ results in the decrease of the quadratic collective error $\gamma_{mean}^2$. Of course, since $\delta$ and $\epsilon_{quad}$ cannot be varied independently of each other,  this interpretation is not correct.  In fact, there is no evidence of meaningful  correlation between the  diversity of the estimates  and the collective error \cite{Reia2021}. We note that $\delta$ is known in the statistical literature as the precision of the estimates, that is, the closeness of repeated estimates (of the same quantity)  to one another \cite{Tan2014}. 

\begin{figure}[t] 
\centering
\includegraphics[width=0.48\textwidth]{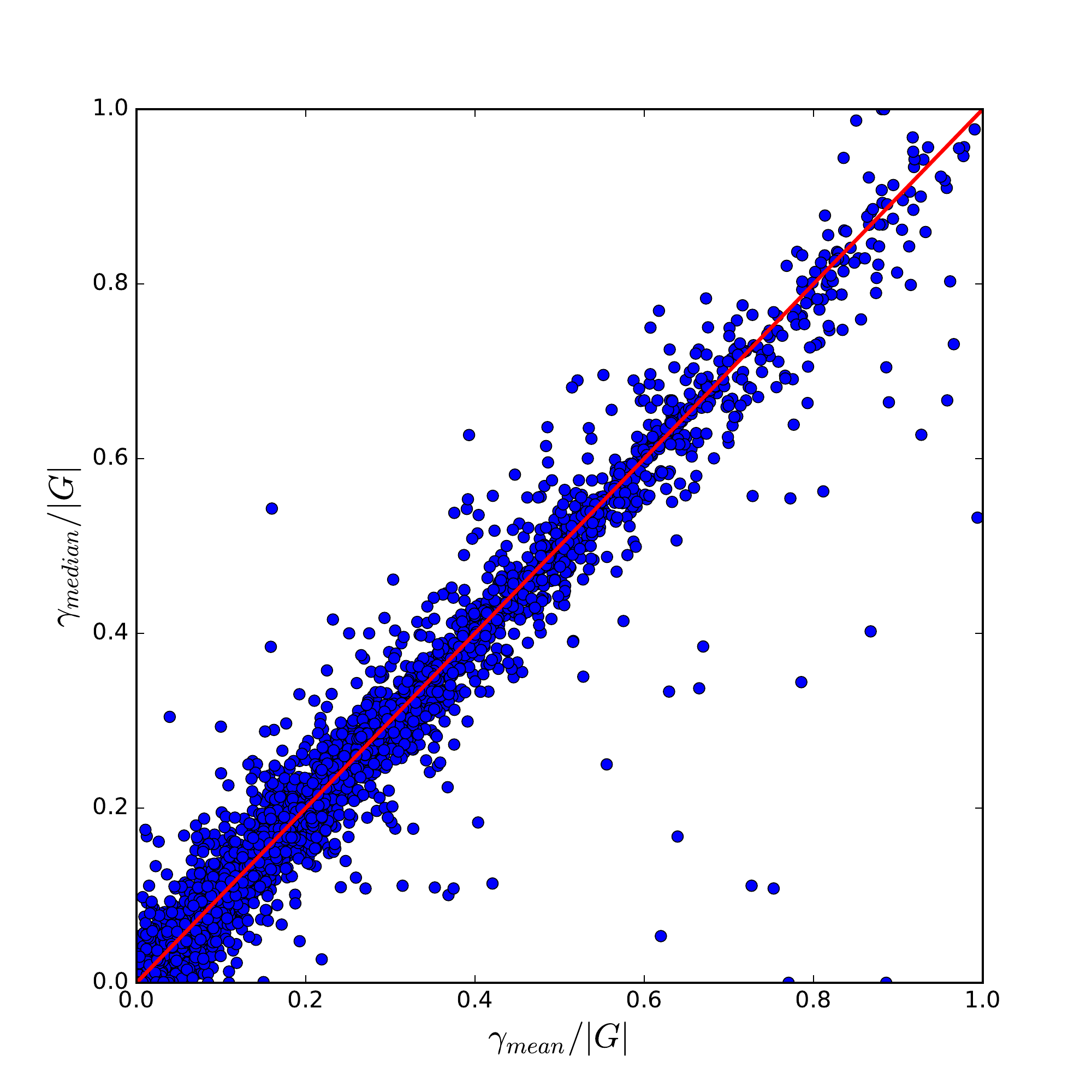}
\caption{Scatter plot of the  relative collective errors   $ \gamma_{mean} /|G|$  and $ \gamma_{median} /|G|$. The solid line is the diagonal  $\gamma_{mean}/|G|  = \gamma_{median}/|G|$.  Pearson's correlation coefficient is $r = 0.98 $. Each symbol stands for one of the $15455$  forecasting contests, but not all contests are shown:    the upper ranges of the $x$ and $y$ axes were limited to 1 to aid visualization, so about 3 percent of the  forecasting contests  are not represented in the figure.
 }  
\label{fig:1}  
\end{figure}

To introduce the median we need to define the empirical  cumulative distribution function \cite{Wasserman2004}
\begin{equation}\label{F1}
 F_n (g) = \frac{1}{n} \sum_{i=1}^n I(g_i \leq g)
\end{equation}
where the indicator function is $I(g_i \leq g) = 1$ if $g_i \leq g$ and $0$ otherwise. The median is then given by 
\begin{equation}\label{med}
 F_n^{-1} (1/2) = \inf \{ g : F_n (g)  > 1/2 \} 
\end{equation}
where $F_n^{-1} (q)$ with $q \in [0,1]$ is the empirical quantile function, i.e., the inverse of the empirical distribution function.  The collective error resulting from the use of the median to aggregate the individuals' estimates is then
\begin{equation}\label{gamma_2}
 \gamma_{median} =  | F_n^{-1} (1/2) - G | .
\end{equation}

Figure \ref{fig:1} shows the scatter plot of the relative collective errors calculated with the mean and the median, as defined in equations (\ref{gamma_1}) and (\ref{gamma_2}), respectively. As expected, there is a high correlation between these two errors, viz., Pearson's correlation coefficient is $r = 0.98 $. The median beats the mean in 50.4 percent  of the  forecasting contests. Note  
 the outliers  with  large  $\gamma_{mean}$ but  small  $\gamma_{median}$ that exemplify  the robustness of the median to extreme individual estimates.  
 
\begin{figure}[t] 
\centering
\includegraphics[width=0.48\textwidth]{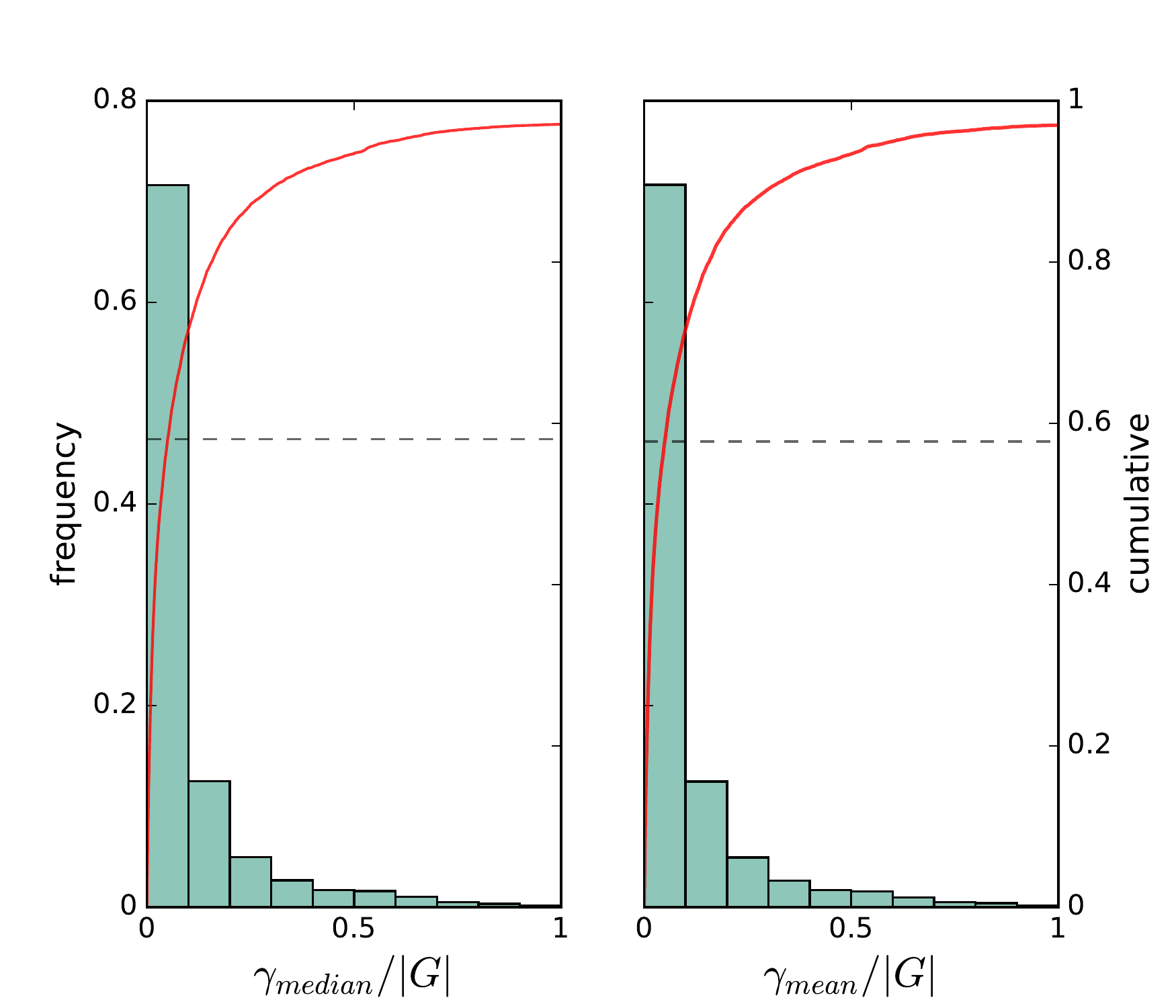}  

\caption{Histograms of the relative collective error calculated with the median $ \gamma_{median} /|G|$ (left panel)  and  of the   relative collective error  calculated with the mean $ \gamma_{mean} /|G|$ (right panel). The red curves are the cumulative distributions and the  horizontal dashed  lines indicate the value of the cumulative distribution when the relative collective error is $ 0.05$. 
 }  
\label{fig:2}  
\end{figure}

To offer more quantitative information on the data shown in the scatter plot, Fig.\ \ref{fig:2} exhibits the histograms and the cumulative distributions of  the number of forecasting contests with a given  relative collective error. The histograms are practically identical for the two aggregation methods,  median and mean. In fact, both methods yield relative   errors less than $0.05$ in 58 percent of the forecasting contests, but their relative errors are greater than 0.1 in  28 percent of the  forecasting contests. In addition,  the mean of the  relative collective error calculated with the mean is $0.20 \pm 0.01$ and the mean of the relative collective error calculated with the median is $0.19 \pm 0.01$. The cumulative distributions exhibited in this figure  show that about 3  percent of the forecasting contests  have  relative collective  errors greater than 1 and are left out of the histograms (as well as of the scatter plot of Fig.\ \ref{fig:1}) for the sake of  visual clarity. We stress, however, that the forecasting contests left out of Figs.\   \ref{fig:1} and  \ref{fig:2}  are not influential in leading to the better overall results for the median.)  In sum, whereas the crowd performance is clearly distinguishable  since 58 percent of its predictions yielded an error  of 5 percent or less, it falls short of being miraculous since  it errs by more than 10 percent in 28 percent of the  forecasting contests. In view of these findings, it seems  likely that selective attention has played a considerable role to secure the popularity of the wisdom of crowds.

\begin{figure}[t] 
\centering
\includegraphics[width=0.48\textwidth]{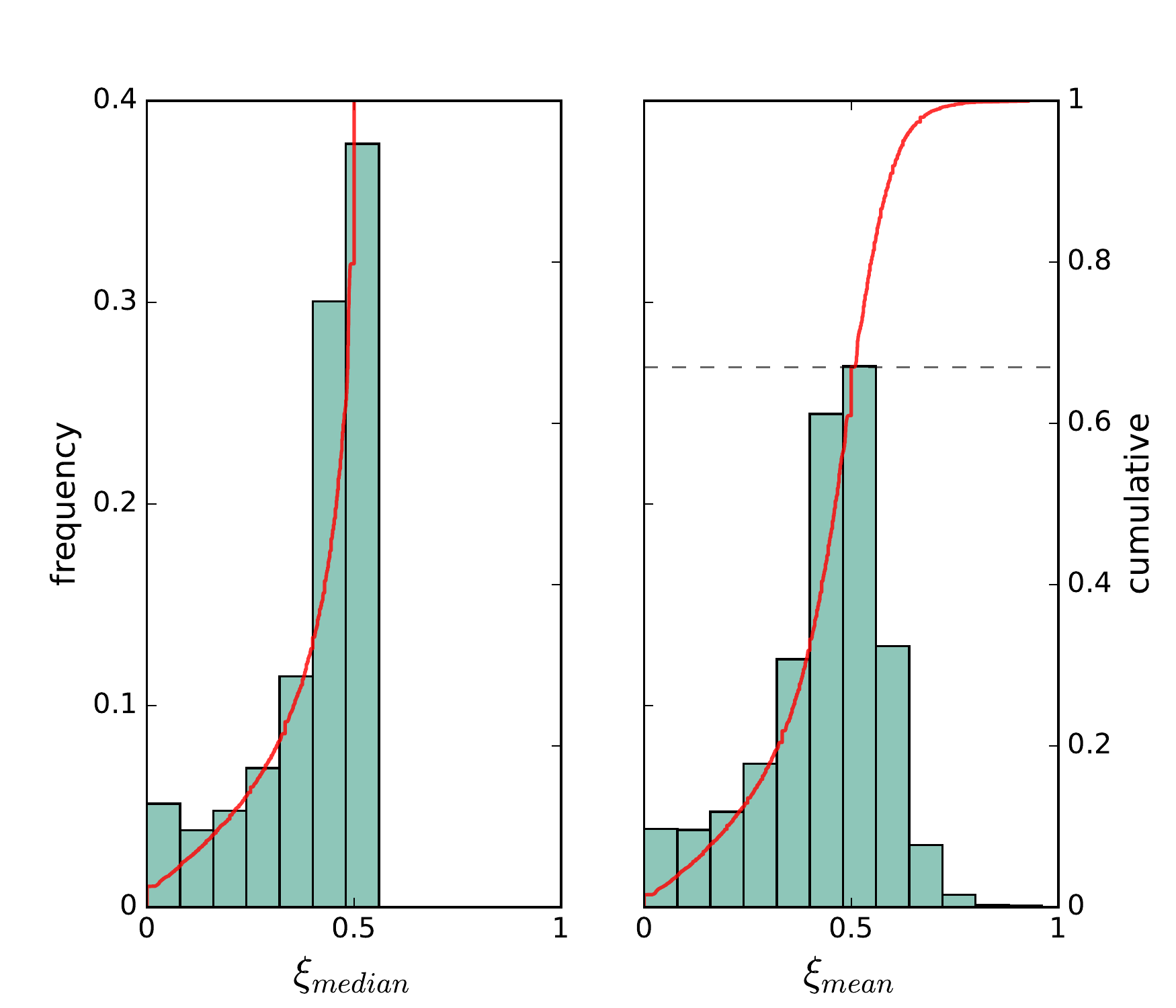}  

\caption{Histograms of the proportion of forecasting contests for which  a fraction $\xi$ of the individuals' estimates  are more  accurate than the relative collective estimate calculated with median  (left panel) and the  relative collective estimate calculated with the mean (right panel). The red curves are the cumulative distributions and the  horizontal dashed  line in the right panel  indicates the value of the cumulative distribution at $\xi_{mean} = 0.5$. 
 }  
\label{fig:3}  
\end{figure}

Next, we measure   the fraction  of individuals' estimates that beat the collective estimate for each forecasting contest. The results are presented in form of histograms and cumulative distributions in Fig.\ \ref{fig:3}, where  the height of the bars is the proportion of forecasting contests for  which that fraction equals $\xi \in \left [ 0, 1 \right ]$. As already  pointed out, the median estimate is always better than the estimates of half of the participants at least, so there is no forecasting contest for which $\xi_{median}>0.5$.  However, the mean    estimate beats the majority of the participants in 67 percent of the forecasting contests only. This number is obtained by evaluating the cumulative distribution function at $\xi_{mean}=0.5$. 
The fraction of forecasting contests for which  $\xi_{mean} =0$ is $0.015$ and the fraction for which  $\xi_{median} =0$ is $0.026$. This is the situation  where the crowd beats all  participants.   Therefore,  if  the  performance criterion is beating most or all participants then the median should definitely  be preferred over the mean as an aggregation method since it is by construction guaranteed to beat at least half  of the participants and has almost double the chances of beating all them as compared with the mean.

%
\section{Testing the unbiased estimates assumption}\label{sec:bias} 
%
As pointed out, a  rather natural explanation for the wisdom of crowds assumes that the  individuals' estimates are unbiased, that is, that the errors spread in equal proportion around the correct value of the unknown quantity. Amusingly,
if  this assumption were correct  one might harvest the benefits of the  wisdom of crowds by asking a single individual to make several  estimates at different times (see, e.g.,  \cite{Vul2008}). Of course, the almost insurmountable difficulty to ensure that a same individual produces a large number of  independent  estimates of a same quantity makes the direct validation  of the unbiased estimates assumption unfeasible. However, we can perform a much simpler and more informative test, viz., to check whether there is a meaningful  correlation between the (unsigned) asymmetry of the distribution of individuals' estimates and the collective error. In fact, according to  the unbiased estimates assumption,  the  more asymmetric  the distribution of estimates is, the more unlikely that  the errors cancel each other  out and hence the greater the collective error.

\begin{figure}[t] 
\centering 
\includegraphics[width=0.48\textwidth]{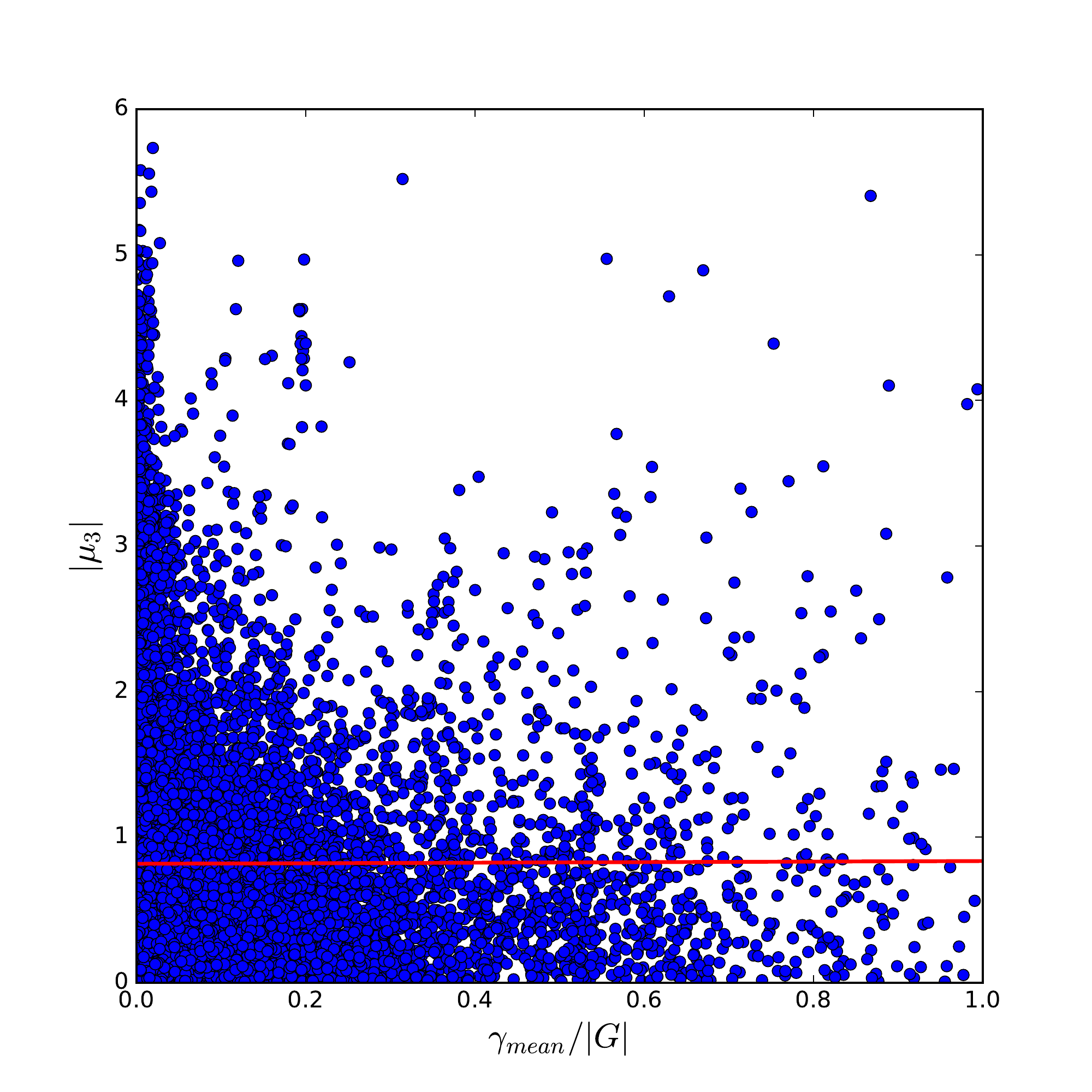}
\caption{Scatter plot of the  relative collective error   $ \gamma_{mean} /|G|$  calculated with the mean and the unsigned  skewness $|\mu_3|$.  The solid line is the linear regression.  The Pearson correlation coefficient is $r = 0.03$.  The upper range of the $x$  axis was limited to 1 to aid visualization, but about  3 percent of the  forecasting contests have relative errors greater than 1.
 }  
\label{fig:4}  
\end{figure}

We recall that the skewness $\mu_3$ of a distribution is  a  dimensionless measure of its asymmetry defined as
\begin{equation}\label{skew}
\mu_3 =  \frac{1}{n} \sum_{i=1}^n \left (\frac{ g_i - \langle g \rangle_n}{\delta^{1/2}} \right )^3,
\end{equation}
where $\langle g \rangle_n$ is the mean of the individuals' estimates defined in equation (\ref{mean}) and $\delta$ is the sample variance  of those estimates defined in equation (\ref{delta}).  A negative value of $\mu_3$ implies that the left tail of the distribution of individual estimates is longer than the right tail, whereas a positive $\mu_3$  indicates a  right-tailed distribution. Since the skewness $\mu_3$ measures the asymmetry with respect to the mean, in this analysis we will only consider  the case that the collective estimate is calculated using  the mean of the individuals' estimates.  

\begin{figure}[t] 
\centering
\includegraphics[width=0.48\textwidth]{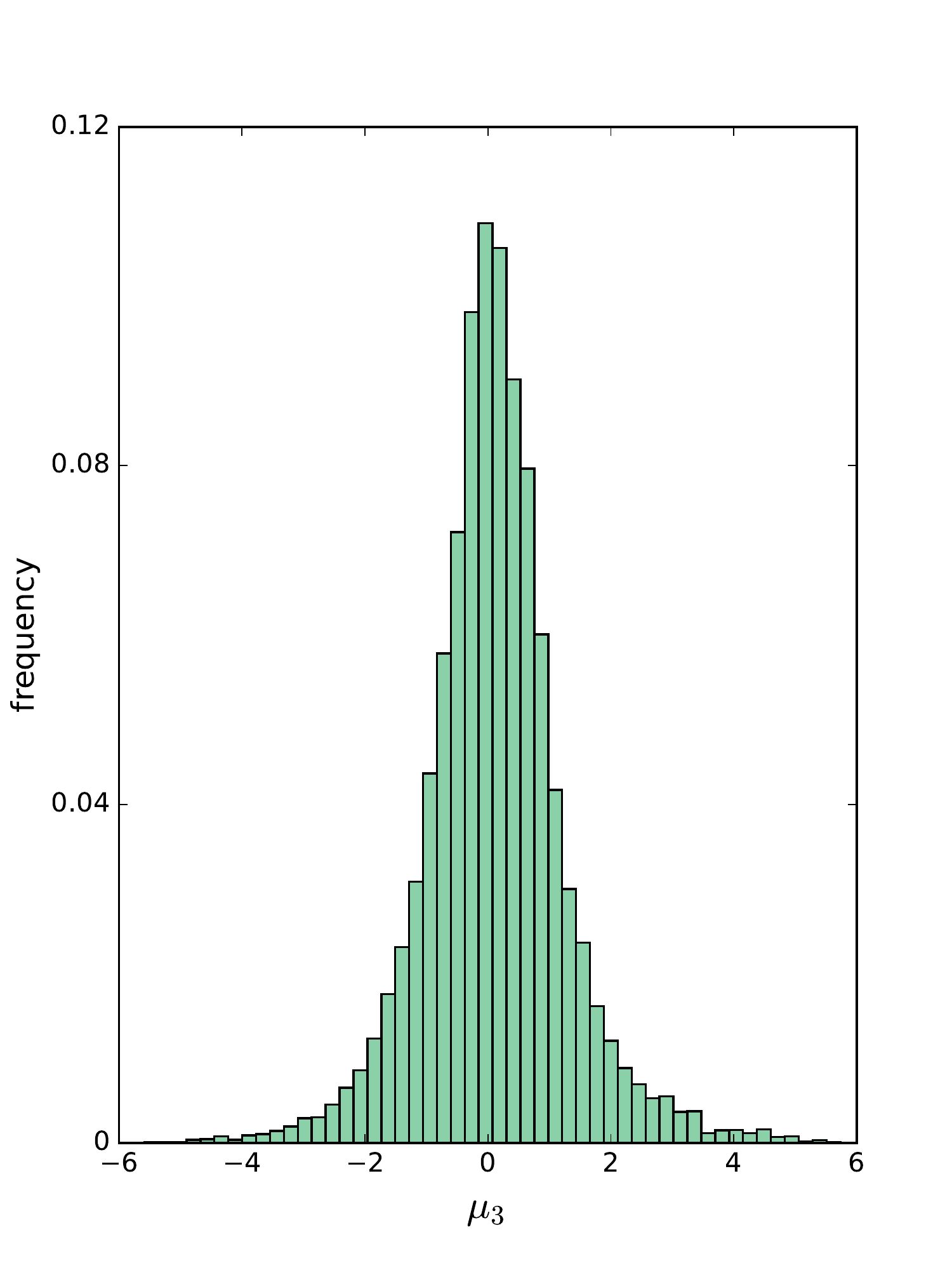}  
\caption{Histogram of the skewness $\mu_3$ of the  distributions of individuals' estimates.  The wisdom of crowds experiments  produce left- and right-tailed distributions in approximately the same proportion.
 }  
\label{fig:5}  
\end{figure}

In Fig.\ \ref{fig:4} we show the scatter plot of the  relative collective error calculated with the mean  and the unsigned  skewness. The results clearly contradict the unbiased estimates assumption, since they indicate that
there is no meaningful correlation between the asymmetry of the  distribution of individuals' estimates and the collective error. More pointedly, Pearson's correlation coefficient for these two variables is $r = 0.03$ and the linear regression is  $|\mu_3| \approx 0.81$. Particularly conspicuous are  the great number of forecasting contests  with high asymmetry and low prediction error, as well as with low asymmetry and high prediction error,  that produce the   triangle-shaped  spreading of points depicted in Fig.\ \ref{fig:4}. It is worth to mention here that if one were to expect that the estimates of a particular  individual were unbiased, i.e., that they  fluctuated symmetrically around the truth,  that individual should most certainly be an expert on the matter in question. Hence our claim  that the asymmetry of the distributions of the experts' predictions in  the  FRBP database debunks the unbiased estimates assumption as an explanation for the accuracy of the crowd's predictions.

Figure  \ref{fig:5} shows that the distribution of the  skewness is approximately symmetric,  i.e., the  number of  right- and left-tailed  distributions of individuals' estimates  are roughly the same  in the forecasting contests of the FRBP database. Hence the asymmetry of those distributions has no role on the wisdom of crowds phenomenon.  In fact,  a scatter plot of the (signed) skewness against  $ \gamma_{mean} /|G|$ yields an almost perfectly symmetric distribution of points around $\mu_3 =0$ (data not shown).

%
\section{Assessing the accuracy of the  forecasters and the crowd}\label{sec:arima} 
%

\begin{figure}[t] 
\centering 
\includegraphics[width=0.48\textwidth]{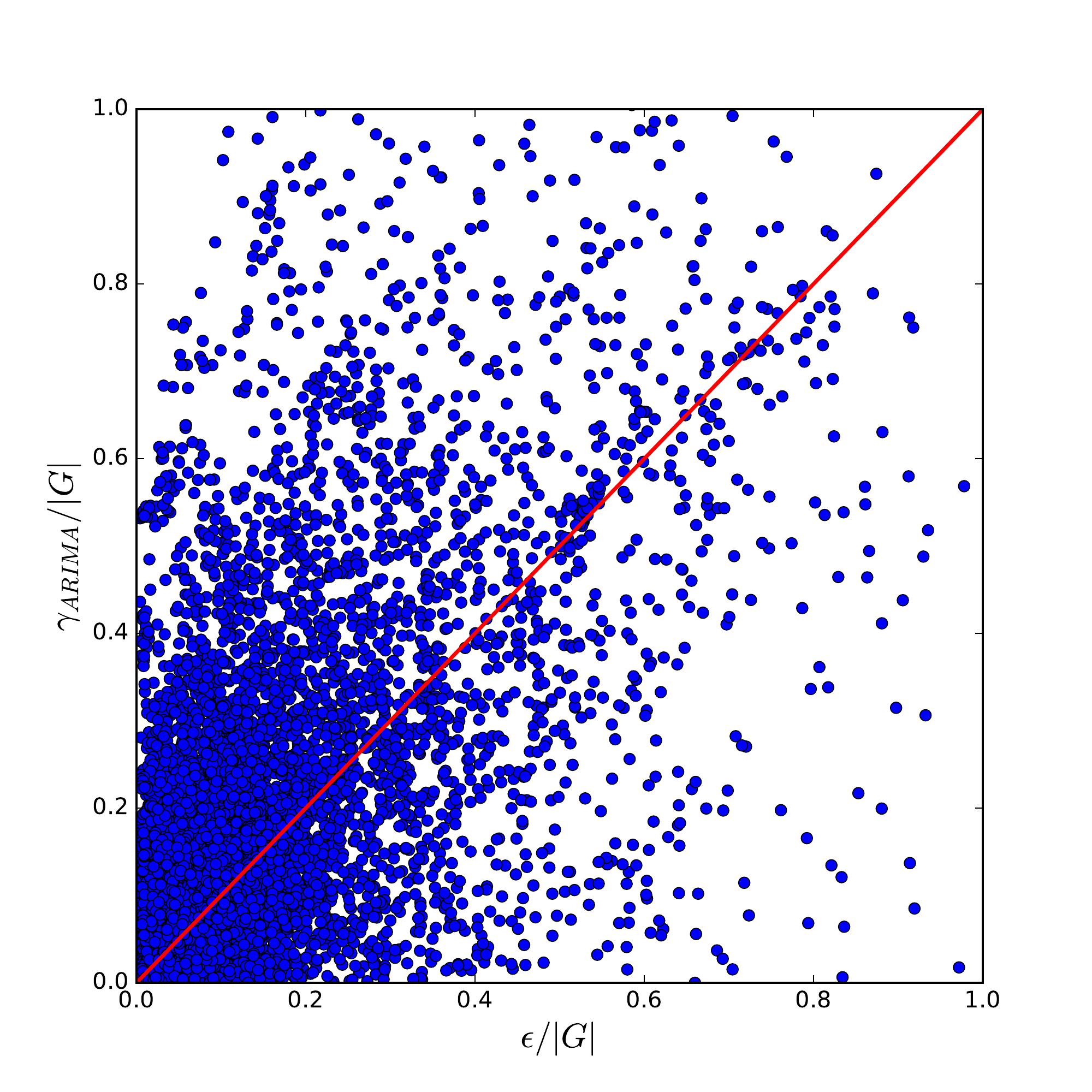}
\caption{Scatter plot of the  relative error of the ARIMA model $\gamma_{ARIMA}/|G|$  and the relative mean individual error $\epsilon/|G|$.  The solid line is the diagonal $\gamma_{ARIMA}/|G| = \epsilon/|G|$.  The upper ranges of the $x$ and $y$  axes were limited to 1 to aid visualization, but about  4 percent of the  forecasting contests have relative errors greater than 1.
 }  
\label{fig:6}  
\end{figure}

Our previous analysis  yielded that the crowd errs by less than 5 percent in 58 percent of the forecasting contests, regardless of the  method - mean or median - used to aggregate  the participants' estimates. Here we address the question: how good  is this performance? Since the crowd beats all participants, which are all expert economists, in less than 3 percent of the  forecasting contests, perhaps an error of 5 percent is nothing to boast about.  Actually, we could as well ask: How good are the expert predictions, anyway? Of course,  to answer those questions we must specify a standard of comparison. The seminal work of Meehl in the 1950s that made the astonishing claim that well-informed experts predicting outcomes were not as accurate as simple algorithms \cite{Meehl1954} (see also \cite{Tetlock2016}) offers a clue.  So here we  use  the basic ARIMA (AutoRegressive Integrated Moving Average) model, which is widely used in  fitting and forecasting nonstationary time series  \cite{Chatfield2000,Hyndman2018}, as the comparison standard to assess the goodness of both forecasters' and  crowd's predictions.

The time series for a fixed variable, say NGDP, is the list of its true values $G$  for each  month since the date the variable was included in the FRBP survey. We used the 12 values  of the variable before the moment of the prediction to set the optimum parameters of the ARIMA  model.  In other words, we train the ARIMA model using a 12-quarters moving window. Once the ARIMA model is trained, we use it to predict the  values of the variable at the five quarters subsequent to the  moment of the prediction, similarly to what is asked to the  participants of the FRBP-SPF.  Since a fraction of the data of the FRBP-SPF database is used to train the ARIMA model, the number of forecasts used in this section is reduced to  $13893$.

Let us first assess the accuracy of the individual forecasters. To achieve that we introduce the mean individual error 
\begin{equation}\label{ind}
\epsilon = \frac{1}{n} \sum_{i=1}^n |  g_i - G | ,
\end {equation}
which can be interpreted as the expected error of an expert selected at random among the $n$ participants of the forecasting contest. Figure \ref{fig:6} shows the scatter plot of the relative error of the ARIMA model $\gamma_{ARIMA}/|G|$ and the relative mean individual error $\epsilon/|G|$ for the
$13893$ forecasting contests. We find that the ARIMA model is more accurate than a randomly selected expert in $35$ percent of the forecasting contests. The average relative mean individual error is $0.22 \pm 0.01$  and the mean relative error of the ARIMA model is $0.31 \pm 0.01$. Hence, at least in economics, the experts have a considerable advantage over  simple time series forecasting algorithms.  

\begin{figure}[t] 
\centering 
\includegraphics[width=0.48\textwidth]{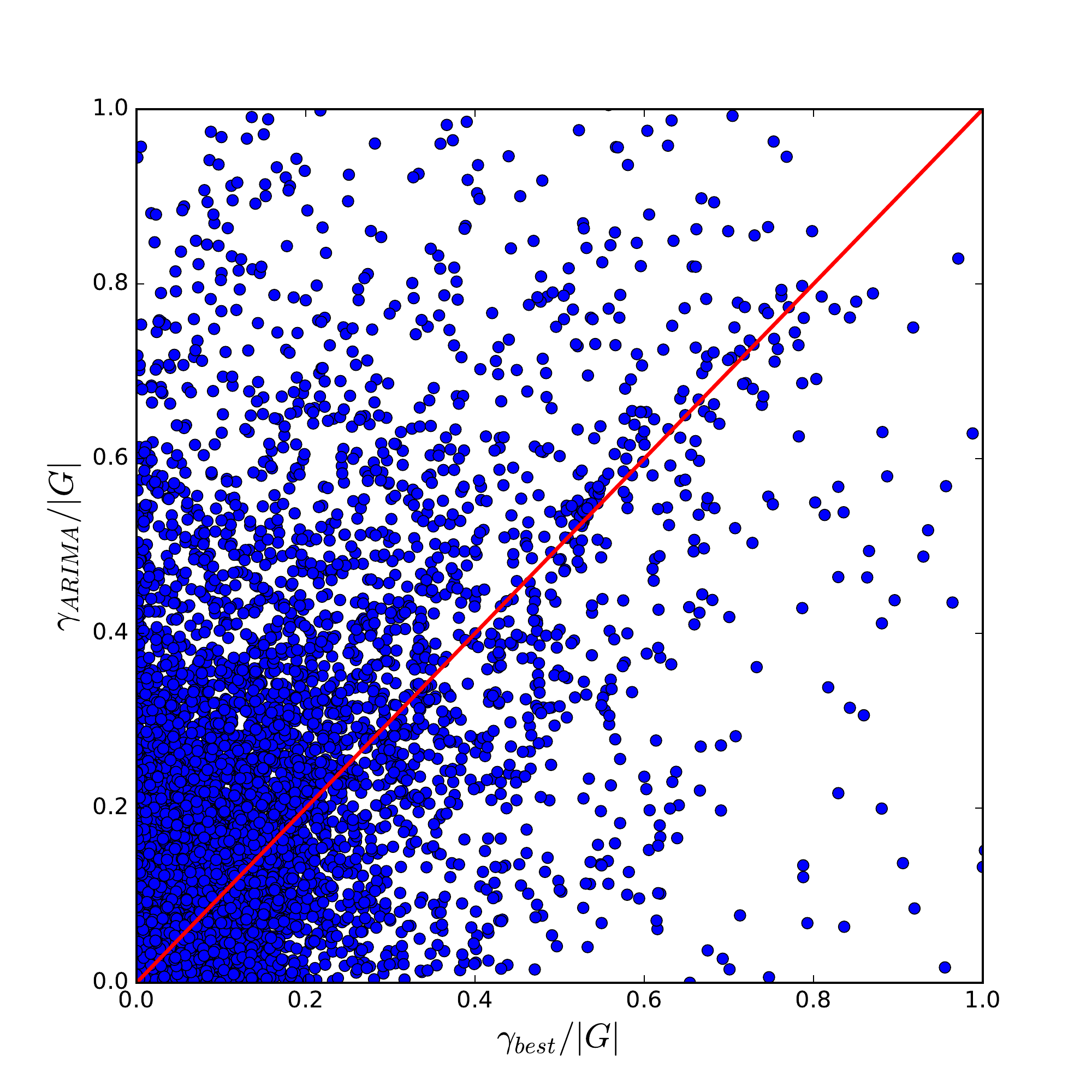}
\caption{Scatter plot of the  relative error of the ARIMA model $\gamma_{ARIMA}/|G|$  and the relative collective error $\gamma_{best}/|G|$.  The solid line is the diagonal $\gamma_{ARIMA}/|G| = \gamma_{best}/|G|$.  The upper ranges of the $x$ and $y$  axes were limited to 1 to aid visualization, but about  4 percent of the  forecasting contests have relative errors greater than 1.
 }  
\label{fig:7}  
\end{figure}

Next we compare the performance of the ARIMA model with  the crowd's performance, which for this analysis we define as $\gamma_{best} = \min \{  \gamma_{mean}, \gamma_{median} \} $.   The results are actually insensitive to this definition because the accuracies of the median and the mean are very close (see Figs.\ \ref{fig:1} and \ref{fig:2}). Figure \ref{fig:7} shows the scatter plot of the relative error of the ARIMA model $\gamma_{ARIMA}/|G|$ and the relative collective error $\gamma_{best}/|G|$. The ARIMA model  beats the crowd  in 28 percent of the forecasting contests.   We recall that the mean relative error of the crowd is $0.19 \pm 0.01$ and the mean relative error of the ARIMA model is $0.31 \pm 0.01$.

An alternative way to assess the goodness of the  crowd's predictions   is to compare them with the  individual predictions  that were closest to the true values, viz., the winning predictions of the forecasting contests. We have already pointed out that the odds that the crowd beats the winner is $0.015$ when  the aggregate of the individuals' estimates is the mean and  $0.026$ when it is the median. Here we add that the mean relative error of the winners is $0.15 \pm 0.01$, so the crowd performance as measured by its mean relative error is really not too far from the best the experts can offer.

%
\section{Conclusion}\label{sec:conc} 
%

The notion that  a group of cooperating individuals  can solve problems more efficiently than  when those same individuals work in isolation is hardly contentious \cite{Rendell2010},   although the factors that make cooperation effective still need some straightening out \cite{Fontanari2014,Reia2019}. In fact,   cooperation may well lead the group astray resulting in the so-called  madness of crowds, as pointed out by MacKay almost two centuries ago \cite{MacKay1841}: ``Men, it has been well said, think in herds; it will be seen that they go mad in herds, while they only recover their senses slowly, and one by one''.  Less dramatically,  cooperation may simply undermine  the benefits of combining independent  forecasts, which is known as the wisdom of crowds \cite{Lorenz2011,King2011}.

More precisely, the wisdom of crowds is  the idea that  the aggregation of the predictions  of many independently deciding individuals is likely to be more accurate than the  predictions  of most  or even all  individuals within the crowd  \cite{Surowiecki2004}.  Here we use the  Federal Reserve Bank of Philadelphia's   Survey of Professional Forecasters (FRBP-SPF) to offer an extensive statistical appraisal of the  wisdom of crowds  for the prediction of  economic indicators.  This approach is immune to the  selective attention fallacy that has most likely influenced the anecdotal evidences for the wisdom of crowds that abound in the literature  \cite{Nobre2020}. In addition, by focusing on the predictions of experts only we are in fact probing the so-called \textit{Vox Expertorum}  \cite{Coste1907}, which is the relevant voice when the forecasts  address economic, social and political issues.

Our analysis of more than ten thousand forecasting contests reveals a  noteworthy  performance of the crowd, but that  can hardly justify the ado about its wisdom. In particular, the odds that the crowd beats all participants of a forecasting contest is $0.015$ when the mean is used to aggregate the individuals' estimates  and $0.026$ when the aggregation is given by the median. This means that the crowd's performance in  Galton's weighing contest, for which the mean collective estimate  has zero error \cite{Wallis2014}, was quite atypical.  The median has clear advantages over the mean as the method to combine the individuals' estimates, specially when the crowd's performance  is compared with  the participants'.  In particular, the median is always guaranteed to beat at least half  of the participants, whereas the mean beats the majority of the participants in 67 percent of the forecasts only (see Fig. \ref{fig:3}) and, as just pointed out, the median has almost double the chances to beat all participants as compared to the mean.  

 Regarding the accuracy of the predictions, however, the advantage of the median is not so noticeable: it is more accurate than the mean in $50.4$ percent of the forecasts. The mean relative errors  of both aggregation methods are  around $20$ percent, which is not really bad considering that the mean relative  error of the winners is $15$ percent and  of a standard prediction algorithm (ARIMA) is $31$ percent. It  is worth mentioning that the expected error of a randomly chosen forecaster is $22$ percent, so the participants of the FRBP survey  predict better than the ARIMA on the average. 

Another significant result of  our extensive statistical analysis  of the  wisdom of crowds is the debunking of the unbiased estimates assumption, which purports to explain the accuracy of the crowd by assuming that the individuals' judgments  scatter symmetrically  around the true value. If this were correct  we should find a strong    correlation between the asymmetry of the distribution of the individuals' estimates and the error of the crowd's estimate.  We find no meaningful correlation between these two quantities (viz., Pearson correlation coefficient is $r =0.03$) and, in fact, there is a large number of  forecasting contests with high asymmetry and low collective error (see Fig. \ref{fig:4}). Because the large number of forecasting contests used in our analysis, the $p$-values associated to the correlation coefficients are extremely small, so all  correlations reported in this paper are statistically significant. Their meaningfulness, however,  is determined by the magnitude of the correlation coefficient $r$. 

Our study of the wisdom of crowds dovetails with  Quetelet's original proposal of Social Physics as an empirical science \cite{Stigler2002,Donnelly2015}, where the regularities of the patterns observed in human behavior and refined through a statistical analysis  are interpreted as the laws of the society (see also \cite{Perc2019}). Perhaps, the statistical  features reported here for the prediction of economic indicators hold true for the prediction of other continuous-valued quantities as well and so they  may shed light on how (expert) humans estimate unknown quantities.

\bigskip

\acknowledgments
The research  of JFF was  supported in part 
 by Grant No.\  2020/03041-3, Fun\-da\-\c{c}\~ao de Amparo \`a Pesquisa do Estado de S\~ao Paulo 
(FAPESP) and  by Grant No.\ 305620/2021-5, Conselho Nacional de Desenvolvimento 
Cient\'{\i}\-fi\-co e Tecnol\'ogico (CNPq).


\begin{thebibliography}{}

\bibitem{Galton1907}
Galton F.  Vox Populi. Nature 1907; 75(1949): 450--451.

\bibitem{Wallis2014}
Wallis KF. 
Revisiting Francis Galton's Forecasting Competition.
Statist Sci. 2014; 29(3):   420--424

\bibitem{Surowiecki2004}
Surowiecki J.  
The Wisdom of Crowds: Why the Many Are Smarter than the Few and How Collective Wisdom Shapes Business, Economies, Societies, and Nations. Doubleday;  2004.

\bibitem{Sunstein2006}
Sunstein  C. 
Infotopia: How Many Minds Produce Knowledge.
Oxford University Press; 2006.

 \bibitem{Page2007}
 Page SE.  The Difference: How the Power of Diversity Creates Better Groups, Firms, Schools, and Societies.
  Princeton University Press; 2007.
   
 \bibitem{Lorenz2011}
Lorenz J, Rauhut H, Schweitzer F,  Helbing D.
How social influence can undermine the wisdom of crowd effect.
Proc Natl Acad Sci USA.  2011; 108(22):  9020--9025.

\bibitem{Bates1969}
 Bates JM,   Granger CWJ.  The combination of forecasts.
 Oper  Res  Q. 1969;   20(4): 451--468.

 \bibitem{Nash2014}
Nash UW.  The Curious Anomaly of Skewed Judgment Distributions and Systematic Error in the Wisdom of Crowds.
 PLoS ONE 2014; 9(11): e112386.
 
 \bibitem{Nash2017}
 Nash UW.  Sequential sampling, magnitude estimation, and the wisdom of crowds. J. Math. Psychol.  2017; 77: 165--179.
 
\bibitem{Reia2021}
Reia SM, Fontanari JF.  Wisdom of crowds: much ado about nothing. J  Stat  Mech. 2021; 2021: 053402.

\bibitem{Nobre2020}
Nobre DA, Fontanari JF. Prediction diversity and selective attention in the wisdom of crowds.  Complex Syst. 2020; 29(4): 861--875.

 \bibitem{FRBP}
 Federal Reserve Bank of Philadelphia.
https://www.philadelphiafed.org/research-and-data/real-time-center/survey-of-professional-forecasters; 2022.

 \bibitem{Clements2022}
Clements MP, Rich RW, Tracy JS.  Surveys of Professionals. Working Paper No. 22-13. Federal Reserve Bank of Cleveland; 2022. https://doi.org/10.26509/frbc-wp-2022-13.

\bibitem{King2011}
King AJ, Cheng L, Starke SD,  Myatt JP.
Is the true `wisdom of the crowd' to copy successful individuals?  Biol Lett.  2011; 8(2): 197--200.
  
\bibitem{Coste1907}
Perry-Coste FH. 
 The ballot-box
 Nature 1907;  75(1952):  509. 
 
 \bibitem{Tan2014}
Tan P-N, Steinbach M,  Kumar V.
Introduction to Data Mining. 
Pearson Education Limited; 2014.

 \bibitem{Wasserman2004}
 Wasserman L.
All of Statistics: A Concise Course in Statistical Inference. 
Springer; 2004.

\bibitem{Vul2008}
Vul E, Pashler H. 
Measuring the crowd within: Probabilistic representations within individuals.
Psychol Sci. 2008; 19(7): 645--647.

\bibitem{Meehl1954}
Meehl P. 
Clinical Versus Statistical Prediction.
University of Minnesota Pres; 1954.

\bibitem{Tetlock2016}
Tetlock PE, Gardner D.
Superforecasting: The Art and Science of Prediction.
Crown Publishing Group; 2016.


\bibitem{Chatfield2000}
Chatfield C. 
Time-series forecasting.
CRC Press; 2000.

\bibitem{Hyndman2018}
Hyndman RJ, Athanasopoulos G.
Forecasting: principles and practice.
OTexts; 2018.

\bibitem{Rendell2010}
Rendell L, Boyd R, Cownden D, Enquist M, Eriksson K, Feldman MW, Fogarty L, Ghirlanda S, Lillicrap T, Laland KN.
Why Copy Others? Insights from the Social Learning Strategies Tournament. Science 2010; 328(5975): 208--213.

\bibitem{Fontanari2014}
Fontanari JF.
Imitative Learning as a Connector of Collective Brains.
PLoS ONE 2014; 9(10):   e110517.

\bibitem{Reia2019}
Reia SM,  Amado AC,  Fontanari  JF.
Agent-based models of collective intelligence.
Phys. Life Rev. 2019; 31: 320--331.

\bibitem{MacKay1841}
MacKay C. 
Extraordinary Delusions and the Madness of Crowds.
Richard Bentley; 1841.

\bibitem{Stigler2002}
Stigler SM.
Statistics on the Table: The History of Statistical Concepts and Methods: The History of Statistical Concepts and Methods.
Harvard University Press; 2002.

\bibitem{Donnelly2015}
Donnelly K.
Adolphe Quetelet, Social Physics and the Average Men of Science, 1796-1874.
University of Pittsburgh Press; 2015.

\bibitem{Perc2019}
Perc M. 
The social physics collective.
Sci. Rep. 2019; 9:16549. 


\end{thebibliography}
\end{document}